\documentclass[journal]{IEEEtran}

\usepackage[dvipsnames]{xcolor}
\usepackage[acronym]{glossaries}
\usepackage[colorlinks, citecolor=blue, linkcolor=blue]{hyperref}
\usepackage{todonotes}
\usepackage{comment}
\usepackage{nameref}
\usepackage{subcaption}
\usepackage{multicol}
\usepackage{multirow}
\usepackage{rotating}
\usepackage{stfloats}
\usepackage[noadjust]{cite}
\usepackage{float}
\usepackage{amsfonts}
\usepackage{upgreek}
\usepackage{mathtools}
\usepackage{scalerel}
\usepackage{lipsum}
\usepackage{boldline}

\newcommand{\SNR}{\text{SNR}}
\newcommand{\SINR}{\text{SINR}}
\newcommand{\dB}{\mathrm{dB}}
\newcommand{\mdg}{\mathrm{mdg}}
\newcommand{\mmse}{\mathrm{mmse}}

\makeatletter
\newcommand*\dotprod{\hspace{2pt}\mathpalette\dotprod@{.5}\hspace{2pt}}
\newcommand*\dotprod@[2]{\mathbin{\vcenter{\hbox{\scalebox{#2}{$\m@th#1\bullet$}}}}}
\makeatother

\let\bigopsize\bigoplus
\def\bigoplus{{\scalerel*{\boldsymbol\oplus}{\bigopsize}}}
\def\bigominus{{\scalerel*{\boldsymbol\ominus}{\bigopsize}}}

\newcommand{\appref}[1]{\hyperref[#1]{appendix~\ref*{#1}}}

\newacronym{lp}{LP}{linearly polarized}
\newacronym{snr}{SNR}{signal-to-noise ratio}
\newacronym{sinr}{SINR}{signal-to-interference-plus-noise ratio}
\newacronym{sdm}{SDM}{space-division multiplexing}
\newacronym{mdg}{MDG}{mode-dependent gain}
\newacronym{mdl}{MDL}{mode-dependent loss}
\newacronym{cdf}{CDF}{cumulative distribution function}
\newacronym{pdf}{PDF}{probability density function}
\newacronym{gue}{GUE}{Gaussian unitary ensemble}
\newacronym{mmse}{MMSE}{minimum mean squared error}
\newacronym{mimo}{MIMO}{multiple-input multiple-output}
\newacronym{msle}{MSLE}{mean squared logarithmic error}
\newacronym{awgn}{AWGN}{additive white Gaussian noise}
\newacronym{nli}{NLI}{non-linear interference}
\newacronym{isi}{ISI}{intersymbol interference}
\newacronym{ml}{ML}{maximum-likelihood}
\newacronym{zf}{ZF}{zero-forcing}
\newacronym{sic}{SIC}{successive interference cancellation}
\newacronym{ssmf}{SSMF}{standard single-mode fiber}
\newacronym{mmf}{MMF}{multi-mode fiber}
\newacronym{fmf}{FMF}{few-mode fiber}
\newacronym{mcf}{MCF}{multi-core fiber}
\newacronym{fec}{FEC}{forward error correction}
\newacronym{ber}{BER}{bit-error rate}
\newacronym{mi}{MI}{mutual information}
\newacronym{air}{AIR}{achievable information rate}
\newacronym{osa}{OSA}{optical spectrum analyzer}
\newacronym{lut}{LUT}{look-up table}
\newacronym{dgd}{DGD}{differential group delay}
\newacronym{fir}{FIR}{finite impulse response}
\newacronym{pmd}{PMD}{polarization-mode dispersion}
\newacronym{csi}{CSI}{channel-state information}
\newacronym{ase}{ASE}{amplified spontaneous emission}
\newacronym{lms}{LMS}{least mean squares}
\newacronym{dsp}{DSP}{digital signal processing}

\begin{document}

\nocite{essiambre2012capacity, morioka2012enhancing, winzer2012optical, richardson2013space, saitoh2016multicore, kahn2017communications, cristiani2022roadmap,winzer2014mode, mello2020impact, shtaif2022challenges}

\title{SDM Optical Systems with MMSE Equalizers: Information Rates and Performance Monitoring}
\author{Lucas~Alves~Zischler,~\IEEEmembership{Graduate~Student~Member,~IEEE,}
        and~Darli~A.~A.~Mello,~\IEEEmembership{Member,~IEEE}%
\thanks{Manuscript received XXX xx, XXXX; revised XXXXX xx, XXXX; accepted XXXX XX, XXXX. This work was financed in part by the Coordenação de Aperfeiçoamento de Pessoal de N\'{i}vel Superior – Brasil (CAPES) – Finance Code 001, CNPq,  and Fapesp grant \#2022/11596-0. (Corresponding author: Lucas Alves Zischler).}%
\thanks{Parts of this work appear in Zischler et al.~\cite{zischler2025analytical}.}%
\thanks{L. Alves Zischler is with the Department of Physical and Chemical Sciences, University of L’Aquila, L’Aquila 67100, Italy: (\mbox{e-mail:~lucas.zischler@univaq.it}).}%
\thanks{D. A. A. Mello is with the School of Electrical and Computer Engineering, State University of Campinas, Campinas 13083-970, Brazil.}}%

\markboth{}{SDM Optical Systems with MMSE Equalizers: Information Rates and Performance Monitoring}%

\maketitle

\begin{abstract}
The information rate of coupled \gls*{sdm} transmission systems is impaired by the stochastic effects of \gls*{mdg} and \gls*{mdl}, turning it into a random variable and reducing its average value. In systems operating with \gls*{mmse} equalizers and no \gls*{csi}, co-channel interference further reduces the instantaneous and average information rates. Analytical solutions for the average information rate in \gls*{mdg}- and \gls*{mdl}-impaired systems under strong coupling have been presented in early studies assuming ideal \gls*{ml} equalization. However, a solution encompassing co-channel interference under \gls*{mmse} equalization has not been presented yet. In this work, we derive statistical models for the \gls*{mmse} equalizer coefficients and develop analytical solutions for the post-filtering information rate. We also use these statistical models and analytical solutions to carry out \gls*{mdg} and \gls*{snr} monitoring in coupled \gls*{sdm} systems. The derived analytical solutions and monitoring techniques are validated by Monte-Carlo simulations, exhibiting a suitable accuracy within practical operational values.
\end{abstract}

\begin{IEEEkeywords}
Space-division multiplexing, MIMO MMSE equalization, information rate, performance monitoring.
\end{IEEEkeywords}

\glsresetall

\section{Introduction}

\IEEEPARstart{N}{ovel} system designs for \gls*{sdm}  transmission have been proposed as a possible step to support the sustained traffic growth in optical networks, particularly in the long-haul and submarine spaces~\cite{essiambre2012capacity, morioka2012enhancing, winzer2012optical, richardson2013space, saitoh2016multicore, kahn2017communications, cristiani2022roadmap}.
In these systems, \glspl*{fmf} and \glspl*{mcf} use multiple spatial channels to increase throughput. However, in coupled propagation schemes, such as those employing \glspl*{fmf} or coupled-core \glspl*{mcf}, deviations in  per-mode\footnote{In this work, the term ''mode`` is used to refer to any supported orthogonal spatial or polarization channel, such as \gls*{lp} modes, polarizations, or cores.} gains and attenuations result in the effects of \gls*{mdg} and \gls*{mdl}\footnote{\Gls*{mdg} and \gls*{mdl} effects are addressed by the same mathematical modeling. Therefore, in this work, we refer to both phenomena as simply \gls*{mdg}.}. In the presence of random mode coupling, these effects become stochastic and reduce average information rates~\cite{winzer2014mode, mello2020impact, shtaif2022challenges}.

Coupled \gls*{sdm} systems require \gls{mimo} receivers to unravel the transmitted signals that are mixed during transmission. These receivers typically implement \gls*{mmse} equalizers owing to their suitable convergence properties, simple feed-forward implementation, and low complexity. However, \gls*{mimo} equalizers operating under the \gls*{mmse} criteria further penalize the information rate of MDG-impaired systems due to unmitigated co-channel interference~\cite{mckay2009achievable}. 
While analytical solutions for the information rates of \gls*{mdg}-impaired systems under ideal \gls*{ml} equalization have been presented in~\cite[Eq.~(10)]{ho2011mode} and~\cite[Eq.~(12)]{mello2020impact}, models accounting for the co-channel interference generated in \gls*{mmse} equalization are still an open problem.

In this paper, we extend~\cite{zischler2025analytical} and derive in detail an analytical expression for the information rate of coupled \gls*{sdm} systems under \gls*{mmse} equalization. The derivations are based on the works by Ho and Kahn in~\cite{ho2011statistics} and~\cite{ho2011mode}, which show that, in \gls*{sdm} systems with strong coupling, the group delay and modal gains in logarithmic scale follow fixed-trace \gls*{gue} random matrices. \Gls*{gue} random matrices have been extensively studied in other fields of physics and mathematics~\cite{wigner1955characteristic, dyson1962statistical, rosenzweig1963graphical, mehta2004random}. The derived expression is exact at the limiting case of an infinite number of supported modes. For a practical number of mode counts, the model provides a valid approximation for average information rate metrics. We apply the analytical method to estimate capacity, pre-\gls*{fec} \gls*{ber}, and constrained capacity considering particular modulation formats. Following the approach in \cite{ho2011mode,ho2013linear,winzer2011mimo,antonelli2015modeling}, and for the sake of clarity, we neglect the effect of nonlinearities, setting an upper bound on capacity. Linear propagation is particularly relevant for current~\gls*{sdm} submarine systems that are not limited by nonlinearities but by power-feed constraints. For further developments on the nonlinearity modeling in~\gls*{sdm} systems, please see~\cite{serena2022ergodic,antonelli2015modelingB}.

This work is structured as follows. Section~\ref{sec:mmse} briefly reviews \gls*{mimo} \gls*{mmse} equalizers, their relation to the optical channel, and the statistics of the channel power gains. Section~\ref{sec:post} presents the analytical solution for the post-filtering \gls*{sinr} as a function of the pre-filtering \gls*{snr} and \gls*{mdg}, and relates the post-filtering \gls*{sinr} to information rate metrics assuming an \gls*{awgn} channel.  Section~\ref{sec:parameter} associates the power gain statistics to the \gls*{mimo} \gls*{mmse} equalizer inverse eigenvalues, and uses the derived analytical formulas for performance monitoring and parameter estimation. Lastly, Section~\ref{sec:conclusion} concludes the paper.

\section{MIMO MMSE Equalization Fundamentals}
\label{sec:mmse}
\subsection{Equalizer matrix}
In coupled \gls{sdm} links, the optical channel operating in the linear domain can be modeled as a frequency-dependent transfer matrix $\mathbf{H}(\omega)\in\mathbb{C}^{D\times D}$, where $D$ is the number of supported polarization and spatial modes. In the derivations, as we calculate averages, we omit the frequency dependence of $\mathbf{H}$ without lack of generality. The considered channel model assumes that the noise is additive and can be accounted independently of the modal coupling matrix $\mathbf{H}$. Under such consideration, we can replace all distributed additive noise sources with an equivalent additive source at the receiver side.

As the transmitted signals couple during propagation, a \gls*{mimo} equalizer is required at the receiver to orthogonalize the incoming streams and recover the transmitted data. Practical \gls*{mimo} equalizers typically employ the \gls*{mmse} criteria owing to the low implementation complexity of \gls*{lms} filters.

The coefficients of the \gls{mimo} \gls{mmse} equalizer, for a particular spectral component, $\mathbf{W}\in\mathbb{C}^{D\times D}$, relate to the channel transfer matrix, $\mathbf{H}\in\mathbb{C}^{D\times D}$, by~\cite[Eq.~(6.44)]{jankiraman2004space}
\begin{equation}
    \mathbf{W}=\left(\mathbf{H}^{H}\mathbf{H}+\frac{\mathbf{I}}{\SNR}\right)^{-1}\mathbf{H}^{H},
\end{equation}
where $\SNR$ is the ratio between total pre-filtering signal power and total \gls*{awgn} power over all $D$ modes. In a high \gls*{snr} scenario, the equalizer matrix approaches the channel matrix inverse (${\lim_{\SNR\rightarrow\infty}\mathbf{W}=\mathbf{H}^{-1}}$).

In coupled \gls*{sdm} systems, the \gls*{snr} accepts several different definitions. In general, in-line \gls*{sdm} amplifiers are usually designed to have a controlled gain, defined by the ratio of their input and output powers. In this operation regime, the amplifier is unable to compensate for losses in individual modes, but a constant total power profile is achieved~\cite{antonelli2015modeling}. We further consider the noise at the receiver to be spatially and spectrally white, leading to equal noise powers across all modes \cite{ho2011statistics}. This condition becomes increasingly accurate for higher mode counts, longer distances, and wider bandwidths~\cite{efimov2014spatial}. Based on these assumptions, we can define the \gls*{snr} as the ratio between total signal optical power to the total optical noise power considering all supported modes
\begin{equation}
    \SNR = \frac{\sum_{i=1}^{D}\sigma^{2}_{u,i}}{D\sigma^{2}_{\eta}},
    \label{eq:snr}
\end{equation}
where $\sigma^{2}_{u,i}$ is the $i^{th}$ mode power and $\sigma^{2}_{\eta}$ is the per-mode noise power.

 In practical systems, optical \gls*{ase} is not the only source of \gls*{awgn}. Transceiver imperfections, such as those generated at the analog-to-digital converter,
 also impose noise contributions that limit the system performance in scenarios of low \gls*{ase}. Furthermore, the excess error generated in equalizers, particularly under the long differential mode delays encountered in SDM systems, also creates an additional noise contribution that affects the source separation process \cite{ho2011statistics, antonelli2015delay, shibahara2017advanced}. The \gls*{snr} referenced in this paper pertains to an equivalent value at the input of the source separation process, achieved by a system free of \gls*{mdg}. Such an approach is in line with~\cite{galdino2017limits}, where the transceiver implementation penalty of transceivers can be measured beforehand in a back-to-back setup and modeled as an \gls*{awgn} source.

The equivalent pre-filtering \gls*{snr} is given by
\begin{equation}
    \SNR=\left(\frac{B_{s}}{12.5\cdot 10^{9}\cdot\text{GSNR}}+\frac{1}{\SNR_{\text{imp}}}\right)^{-1},
    \label{eq:snrimp}
\end{equation}
where $\text{GSNR}$ is the measured generalized optical \gls*{snr} in a 12.5~GHz bandwidth, and $B_{s}$ is the signal bandwidth.

The $\SNR_{\text{imp}}$ is measured in a setup with no \gls*{ase} noise or \gls*{mdg}, but with \gls*{dgd} to account for the equalizer excess error generated, e.g., when its length is too short to compensate for the channel delay spread. The characterization of the excess error contribution for $\SNR_{\text{imp}}$ can be carried out by simulation or experimentally using the statistical models presented in~\cite{ho2011statistics}. 
It is important to emphasize that short links, where 
$\SNR_{\text{imp}}$ plays a more significant role, typically exhibit low delay spreads, resulting in minimal equalizer excess errors. On the other hand, in long \gls*{sdm} links, where the \gls*{dgd} is most pronounced, the contribution of  $\SNR_{\text{imp}}$ becomes negligible because of intensive \gls*{ase} accumulation.  

\subsection{Channel Power Gains Statistics}
\label{sec:power}

Assuming a linear \gls*{mimo} \gls*{awgn} channel, the linear scale eigenvalues $\lambda_{i}$ of the $\mathbf{H}\mathbf{H}^{H}$ operator can be interpreted as power gains for the individual channels~\cite{paulraj2003introduction}. The eigenmodes obtained from the eigendecomposition operation are the Schmidt modes of the channel and represents an equivalent optimal basis of propagation~\cite{ho2013linear}. The ratio between the signal power propagated along the $i^{th}$ Schmidt mode and its respective noise power is given by ${\lambda_{i}\cdot\SNR}$. The eigenvalues are sorted in ascending order (${\lambda_{1}\le\lambda_{2}\le\cdots\le\lambda_{D}}$).

As pointed out in~\cite{ho2011mode}, the channel power gains in decibel units follow the well-defined fixed-trace \gls*{gue} spectral \gls*{pdf}. The unlabeled\footnote{The unlabeled distribution considers the ensemble of all modal-gains $\lambda$ regardless of index.} decibel power gain distribution is given by~\cite[Table~1]{ho2011mode},~\cite[Eq.~(S.15)]{ho2011statistics}
\begin{equation}
	\begin{split}
		f_{\lambda_{\dB}}(\lambda_{\dB}) = &\frac{\alpha_{\lambda_{\dB},D}}{\sigma_{\mdg}}e^{-\frac{(D+1)}{2}\left(\frac{\lambda_{\dB}-\mu_{\lambda_{\dB}}}{\sigma_{\mdg}}\right)^{2}}\\
		&\times\sum_{k=0}^{D-1}\beta_{\lambda_{\dB},D,k}\left(\frac{\lambda_{\dB}-\mu_{\lambda_{\dB}}}{\sigma_{\mdg}}\right)^{2k},
	\end{split}
	\label{eq:flambdadb}
\end{equation}
where $\alpha_{\lambda_{\dB},D}$ is a normalization factor, $\beta_{\lambda_{\dB},D,k}$ are polynomial coefficients, $\mu_{\lambda_{\dB}}$ is the average logarithmic gain such that the total linear power gain is unitary, and $\sigma_{\mdg}$ is the standard deviation of the unlabeled logarithmic eigenvalues, which is a common metric in literature to quantify the \gls*{mdg}. The coefficients of~\eqref{eq:flambdadb} are described in further detail in~\cite{ho2011mode} and~\cite{zischler2025analytic}. At the limit $D\rightarrow\infty$,~\eqref{eq:flambdadb} converges to the semicircular distribution presented in ~\cite[Table~1]{ho2011mode},~\cite[I]{wigner1955characteristic}
\begin{equation}
    \begin{gathered}
	    f_{\lambda_{\dB}}(\lambda_{\dB}) = \frac{1}{2\pi\sigma_{\mdg}}\sqrt{4-\frac{\left(\lambda_{\dB}-\mu_{\lambda_{\dB}}\right)^{2}}{\sigma_{\mdg}^{2}}},\\
        -2\sigma_{\mdg}+\mu_{\lambda_{\dB}}\leq\lambda_{\dB}\leq 2\sigma_{\mdg}+\mu_{\lambda_{\dB}}.
    \end{gathered}
    \label{eq:flambdadbwigner}
\end{equation}

\section{Information Rates in MIMO Systems Using MMSE Receivers}
\label{sec:post}

Optical \gls*{sdm} channels with \gls*{mdg} and random coupling exhibit channel gains $\lambda_i$ that vary over time but preserve $\sum_{i=1}^D \mathbb{E}\{\lambda_i\} = D$. As a consequence, the average per-mode channel gain $\mathbb{E}\{\lambda\}$ is unitary. The information rate losses caused by \gls*{mdg} can be proven for any \gls*{mdg} distribution from Jensen's inequality, given that the logarithmic relation between capacity and \gls*{snr} is concave~\cite{jensen1906fonctions}. Applying Jensen's inequality for concave functions to the Shannon's capacity formula yields
\begin{equation}
    \mathbb{E}\left\{\log_{2}\left(1+\lambda\cdot\SNR\right)\right\}\leq\log_{2}\left(1+\mathbb{E}\left\{\lambda\right\}\SNR\right),
\end{equation}

As $\mathbb{E}\{\lambda\}=1$, the expected per-mode capacity under \gls*{mdg} is, at most, equal to that of an equivalent single-mode link
\begin{equation}
    \mathbb{E}\left\{\log_{2}\left(1+\lambda\cdot\SNR\right)\right\}\leq\log_{2}\left(1+\SNR\right),
\end{equation}
where the equality holds only if\footnote{${\mathbb{E}\{f(X)\}=f(\mathbb{E}\{X\})}$ only holds if $f(\cdot)$ is affine or $\sigma_{X}=0$. This is a known result of functional inequalities~\cite{grimmett2020probability}. As the logarithmic relation between capacity and \gls*{snr} is strictly concave, only the latter case provides the equality.} $\sigma_{\mdg}=0~\dB$. This \gls*{mdg} impairment appears at the receiver as a \gls*{snr} penalty~\cite{mello2020impact}.

\subsection{Derivation of the post-MMSE filtering SINR}

Several studies on the impact of \gls*{mdg} on the channel capacity assume an optimal \gls*{ml} receiver. However, coupled \gls*{sdm} systems using \gls*{mmse} equalizers have further information rate losses because of co-channel interference~\cite{mckay2009achievable}\footnote{In \gls*{mimo} \gls*{mmse} equalizers parallel data streams generate co-channel interference. More sophisticated equalization techniques, such as \gls*{sic} equalization, approach the performance of ideal \gls*{ml} filtering~\cite{chou2022successive}.}. Therefore, the post-filtering \gls*{snr} in a particular filter output is rather given by a \gls*{sinr}. The \gls*{sinr} in a particular filter output $i$ relates to the channel matrix coefficients by~\cite[Eq.~(9)]{mckay2009achievable}
\begin{equation}
    \SINR_{i}=\frac{1}{\left[\left(\mathbf{I}+\SNR\cdot\mathbf{H}^{H}\mathbf{H}\right)^{-1}\right]_{i,i}}-1.
    \label{eq:sinr}
\end{equation}

Equation ~\eqref{eq:sinr} provides an instantaneous $\SINR_{i}$ for an instantaneous matrix realization $\mathbf{H}$. However, for some applications, such as calculating the average information rates in optical \gls*{sdm} systems, an average $\SINR_{i}$ is also desired. Until now, calculating the average \gls*{sinr} of signals affected by \gls*{mdg} after an \gls*{mmse} equalizer has been accomplished by Monte-Carlo simulation of several realizations of $\mathbf{H}$. As the main contribution in this paper, we show in this section that an approximation for the average \gls*{sinr} in \gls*{mdg}-impaired signals can be analytically computed as
\begin{equation}
    \SINR = \left[\int_{0}^{1}\frac{10\cdot f_{\lambda_{\dB}}\left(10\cdot\log_{10}\left(\frac{1-x}{\SNR\cdot x}\right)\right)}{\ln(10)(1-x)}dx\right]^{-1} - 1.
    \label{eq:sinrint}
\end{equation}
The expression is analytically exact at the limiting case of $D\rightarrow\infty$ \footnote{Throughout the rest of the paper, variable \gls*{sinr}, without index, represents this exact solution.}. The full derivation is presented in~\appref{app:a}.

In the limiting case of $D\rightarrow\infty$, as the standard deviation of the per-mode gains scales by ${1/D}$~\cite[Eq.~(6)]{zischler2025analytic}, the $\lambda_{i}$ values converge to deterministic solutions and the \gls*{sinr} of all modes approach~\eqref{eq:sinrint}. This effect of reduced variance due to increased mode diversity has been previously discussed in~\cite{arik2014diversity,mello2023impact}.

Under finite mode counts, the solution given in~\eqref{eq:sinrint} can be used as an approximation for the average \gls*{sinr}, as
\begin{equation}
    \mathbb{E}\left\{\sum_{i=1}^{D}\frac{1}{D}\SINR_{i}\right\} \approx \SINR.
    \label{eq:sinrappr}
\end{equation}

\begin{figure}[t]
    \centering
    \includegraphics{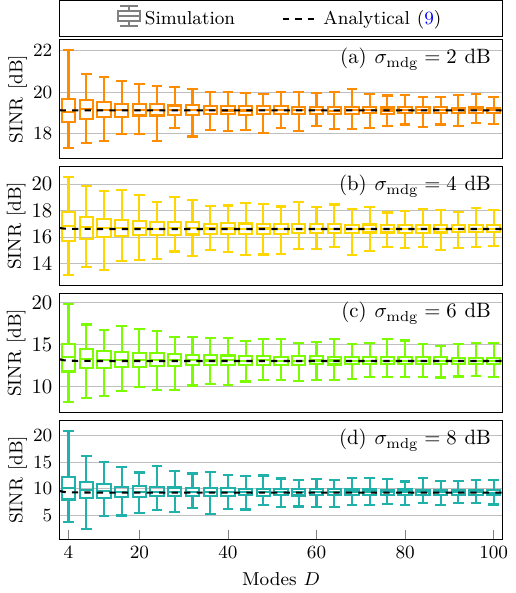}
    \caption{Box plots of the SINR simulated values compared to the analytical solution with respect to the mode-count for $\SNR=20~\dB$. The simulations  rely on the parameters given in Table~\ref{tab:parameters}, with a modification on the frequency bin count (to 1 frequency bin) for no frequency diversity. The simulated distribution considers all modes regardless of the index.}
    \label{fig:SinrVariance}
\end{figure}

\begin{table}[t]
    \centering
    \begin{tabular}{|c|c|} 
         \hline
         \textbf{Parameter} & \textbf{Value} \\\hline
         \parbox[c]{4cm}{\vspace{1mm}\centering Monte-Carlo trials per point} & 20 \\
         Supported modes & \parbox[c]{4cm}{\centering 6 (3 spatial modes$\times$2 polarizations)} \\
         Number of frequency bins & 1000 \\
         Chromatic dispersion & 22 ps/nm/km \\
         Per-section DGD & 3.1 ps/$\sqrt{\text{km}}$ \\
         Number of sections & 100 \\
         Section length & 50 km \\
         Carrier Frequency & 1550 nm \\
         Symbol rate & 30 GBaud \\
         Time domain channel granularity & 8 samples/symbol \\
         RX oversampling rate & 2 samples/symbol \\
         Equalizer taps & 1000 \\
         Training symbols & $8\cdot 10^{5}$ \\
         Total number of symbols & $10\cdot 10^{5}$ \\\hline
    \end{tabular}
    \caption{Simulation parameters used for Monte-Carlo simulations for a single frequency channel.}
    \label{tab:parameters}
\end{table}

Fig.~\ref{fig:SinrVariance} shows box plots for the post-equalization \gls*{sinr}, obtained by simulation. The parameters used in all simulations are given in Table~\ref{tab:parameters}, unless explicitly stated otherwise. To generate Fig.~\ref{fig:SinrVariance}, the channel coefficients $\lambda_i$ are randomly simulated and evaluated through matrix operations. The simulated channel matrices are generated using the multi-section model\footnote{The multi-section model accounts for random modal coupling and MDG, neglecting nonlinearities.} presented in~\cite[Eq.~(2)]{ho2011mode}. The impact of fiber attention appears as a reduced SNR at the end of the link. The simulations for Fig.~\ref{fig:SinrVariance} consider a worst-case scenario of only 1 frequency bin, as frequency diversity reduces the variance of information rate metrics~\cite{ho2011frequency}. Fig.~\ref{fig:SinrVariance} reveals a good agreement between the simulated average \gls*{sinr} and the results generated by the analytical formula, with the \gls*{sinr} variance decreasing towards higher mode counts owing to mode diversity.

\begin{figure*}[!t]
    \centering
    \includegraphics{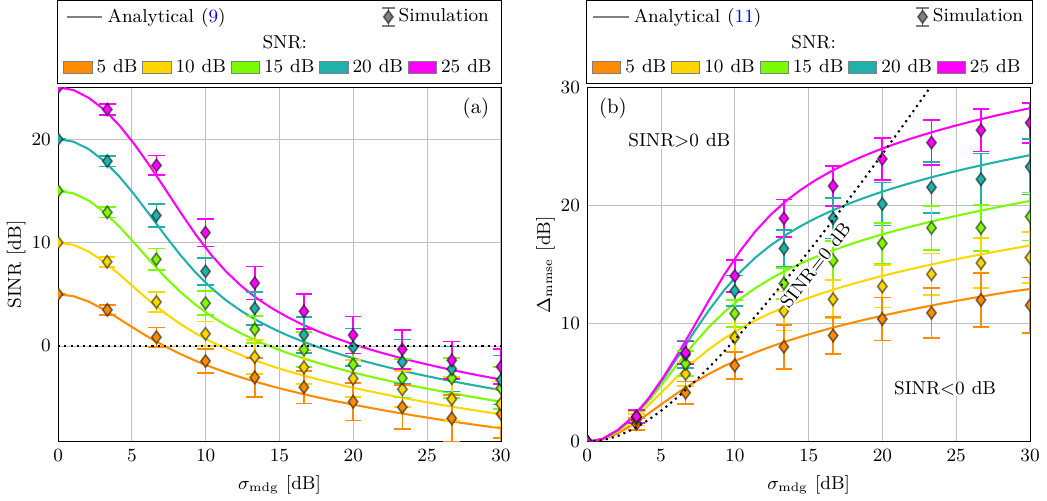}
    \caption{(a) Average SINR as a function of $\sigma_{\mdg}$ for $D=6$. (b) Effective SNR loss due to MDG as a function of $\sigma_{\mdg}$ for $D=6$. The 0~dB SINR line in (b) is obtained from the $\sigma_{\mdg}$ values where ${\Delta_{\mmse}=10\cdot\log_{10}(\SNR)}$. The curves obtained from Monte-Carlo simulation are presented alongside the analytical predictions. The markers show the simulation sample mean, and error bars correspond to one sample standard deviation.}
    \label{fig:Sinr}
\end{figure*}

To assess the approximation in~\eqref{eq:sinrappr}, we compare the analytical solution with simulations of a 3-mode polarization multiplexed link (${D=6}$), which is the lowest mode count supported in \glspl*{fmf}, resulting in the highest divergences to~\eqref{eq:sinrint} because of low mode diversity. The simulation and analytical curves of average \gls*{sinr} are presented in Fig.~\ref{fig:Sinr}(a). The results reveal that~\eqref{eq:sinrint} is a valid approximation for practical values of \gls*{mdg} (${\sigma_{\mdg}<10~\dB}$) even for low mode counts. At high \gls*{snr} and high \gls*{mdg} values, the analytical curve underestimates the average \gls*{sinr} by a small margin.

A more tangible metric to express the impact of \gls*{mdg} on the system performance is the effective \gls*{snr} loss, defined as~\cite[Eq.~(14)]{mello2020impact}
\begin{equation}
    \Delta_{\mmse}=10\cdot\log_{10}(\SNR)-10\cdot\log_{10}(\SINR).
    \label{eq:effsnrloss}
\end{equation}

Fig.~\ref{fig:Sinr}(b) demonstrates severe losses caused by \gls*{mdg}, resulting in several decibels of penalty. In some cases, the \gls*{mdg}-induced penalties can nullify any benefits of low \gls*{awgn} setups. However, practical systems should not be allowed to operate under such extreme \gls*{mdg} conditions. In any case, the analytical model reveals a suitable accuracy in predicting effective \gls*{snr} losses, particularly for ${\sigma_{\mdg}<10~\dB}$.

\subsection{Capacity}
In the previous section, we derived an analytical expression for the signal \gls*{sinr} after \gls*{mmse} filtering. In this section, we apply~\eqref{eq:sinrint} to the Shannon formula to calculate the capacity of \gls*{mimo} channels incorporating \gls*{mmse} equalizers. The average per-mode capacity with \gls*{mmse} equalization and non-zero \gls*{mdg}, under the \gls*{awgn} channel model, in bits/s/Hz/mode, is given by
\begin{equation}
    C_{\mmse} = \frac{1}{D}\sum_{i=1}^{D}\log_{2}(1+\SINR_{i}).
    \label{eq:capmmse}
\end{equation}

As previously discussed,  we use the limiting case of an infinite number of modes to derive an approximation for the average SINR, yielding a limiting value for the per-mode capacity 
\begin{equation}
     C_{\mmse} \approx \lim_{D\rightarrow\infty}\frac{1}{D}\sum_{i=1}^{D}\log_{2}(1+\SINR_{i}) = \log_{2}(1+\SINR).
    \label{eq:caplim}
\end{equation}

\begin{figure*}[!t]
    \centering
    \includegraphics{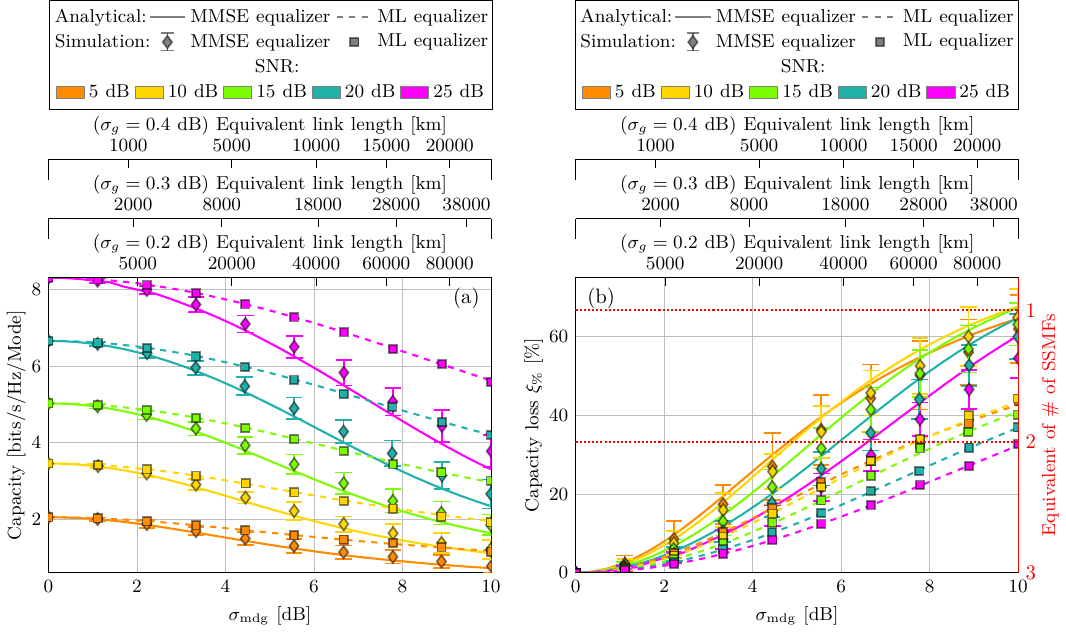}
    \caption{(a) Average capacity curves as a function of $\sigma_{\mdg}$ for $D=6$. (b) Capacity loss due to MDG as a function of $\sigma_{\mdg}$ for $D=6$. The MMSE equalizer analytical curves are derived from~\eqref{eq:caplim}, whereas the ideal ML equalizer curves are derived from~\cite[Eq.~(10)]{ho2011mode}. The curves obtained from Monte-Carlo simulations are presented alongside the analytical predictions. The markers show the simulation sample mean, and error bars correspond to one sample standard deviation. In (b) it is shown, for comparison, the number of SSMFs that yield equivalent capacity. We also present the equivalent transparent reach as a function of $\sigma_{\mdg}$ for selected per-section MDG ($\sigma_{g}$) values, considering a 50-km section length.}
    \label{fig:Capacity}
\end{figure*}

Fig.~\ref{fig:Capacity}(a) compares analytical and simulation values, for $C$ under both ideal \gls*{ml} equalization and \gls*{mmse} equalization. Both capacity values are severely impaired by \gls*{mdg}, with setups applying \gls*{mmse} filtering suffering a higher penalty. The curves reveal excellent agreement between simulation and the analytical model.

As discussed in~\cite{mello2020impact}, capacity loss metrics are also relevant to express the penalties caused by \gls*{mdg}. The capacity loss metric defined in~\cite{mello2020impact} is given by 
\begin{equation}
    \xi = 1-\frac{C_{\mmse}}{C_{\text{awgn}}},
\end{equation}
where $C_{\text{awgn}}$ is the average per-mode capacity of an equivalent link with only \gls*{awgn} and without \gls*{mdg}. In the scope of this paper, $C_{\text{awgn}}$ is defined as
\begin{equation}
    C_{\text{awgn}} = \frac{1}{D}\sum_{i=1}^{D}\log_{2}(1+\SNR) = \log_{2}(1+\SNR).
\end{equation}

The capacity loss evaluated for simulations, and for the analytical model is shown in Fig.~\ref{fig:Capacity}(b), as a function of \gls*{mdg}. As discussed in previous papers, there is a significant capacity reduction if \gls*{mdg} is left unmanaged. These penalties are especially relevant with \gls*{mmse} equalization. The \gls*{mdg} impact can become so significant that the capacity of the spatially multiplexed system can become equivalent to that of a single \gls*{ssmf}, eliminating all the benefits of \gls*{sdm}. 

Fig.~\ref{fig:Capacity} also relates capacity metrics to the span count~\cite{ho2011mode}, highlighting the analytical model as a powerful design tool. The relationship is carried out assuming an optical link with homogeneous spans with constant per-section \gls*{mdg} standard deviation ($\sigma_{g}$). The approximate formula provided in~\cite[Eq.~(1)]{ho2011mode}, which is accurate for $\sigma_{\mdg}$ values up to 10~dB, enables to relate $\sigma_{\mdg}$ and distance for selected per-section $\sigma_{g}$ values. Such transparent reach analysis has been discussed in greater detail in~\cite{mello2020impact} and~\cite{mello2023impact}. Fig.~\ref{fig:Capacity} confirms the large impact of minimal changes in per-section \gls*{mdg} values on capacity in long-haul terrestrial and submarine links with high accumulated \gls*{mdg}~\cite{bruyere1994penalties,pilipetskii2006performance,shtaif2004polarization}.

Due to the stochastic nature of \gls*{mdg}, variations in information rate are expected and can result in outages. For non-infinite mode-count links, the \gls*{sinr} is a stochastic value as seen in Fig.~\ref{fig:SinrVariance}, and the calculation of such variance is still an open problem. For optimal \gls*{ml} equalization, analytical models for the capacity variance are provided in~\cite{zischler2025analytic}. 

\subsection{Pre-FEC BER and Constrained Capacity}
\label{ssec:pre}

Complementing the previous section on capacity evaluation, this section investigates the performance of the proposed analytical model in predicting the pre-\gls*{fec} \gls*{ber} and constrained capacity considering particular modulation formats. In a $M$-QAM modulated transmission scheme with Gray mapping, the \gls*{sinr} relates to the pre-FEC BER by~\cite[Eq.~(6.20)]{grami2015introduction}
\begin{equation}
    b_{e}\approx\frac{2(M-1)}{M\log_{2}(M)}Q\left(\sqrt{\frac{6}{M^{2}-1}\SINR}\right),
    \label{eq:ber}
\end{equation}
where $Q(\cdot)$ is the Q-function.
\begin{figure*}[!t]
    \centering
    \includegraphics{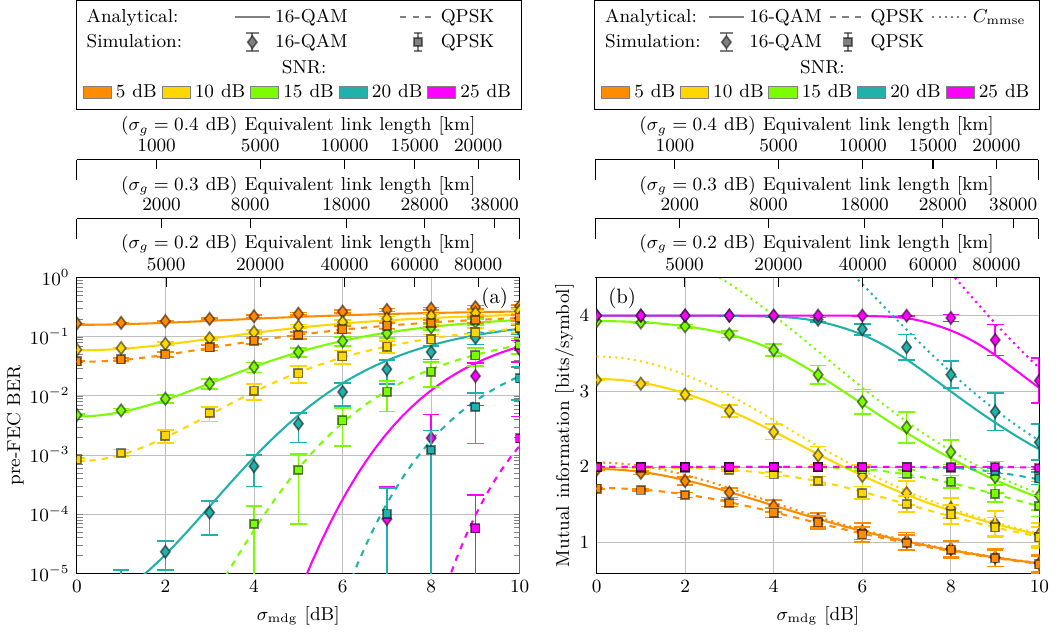}
    \caption{(a) Average pre-FEC BER as a function of $\sigma_{\mdg}$ for $D=6$. (b) MI as a function of $\sigma_{\mdg}$ for $D=6$. The pre-FEC BER is obtained from~\eqref{eq:ber}, whereas the MI is obtained from~\eqref{eq:mi} considering a 10-point Gauss-Hermite quadrature. The curves obtained from Monte-Carlo simulations are presented alongside the analytical results. The markers show the simulation sample mean, and error bars correspond to one sample standard deviation. The simulation considers the transmitter and receiver as given by Table~\ref{tab:parameters}, whereas the BER is obtained from the post-decision binary sequence and the MI is derived from the post-filtering symbol sequence for an AWGN channel, by the method described in~\cite{buchali2015rate}. In (b), we show the capacity alongside the mutual information obtained from~\eqref{eq:caplim}. We also present the equivalent transparent reach as a function of $\sigma_{\mdg}$ for selected per-section MDG ($\sigma_{g}$) values, considering a 50-km section length.}
    \label{fig:InformationRate}
\end{figure*}

In addition to the pre-\gls*{fec} \gls*{ber}, the constrained capacity, given in bits per symbol, is also a relevant metric to incorporate rate limitations imposed by modulation~\cite{fehenberger2015achievable}. The constrained capacity, or \gls*{mi}, corresponding to a $M$-QAM sequence, can be obtained by the $J$-point Gauss-Hermite quadrature approximation given by~\cite[Eq.~(40)]{alvarado2018achievable}
\begin{equation}
    \begin{split}
        I\approx&\log_{2}(M)-\frac{1}{M\pi}\sum_{(i,l_{1},l_{2})=(1,1,1)}^{(M,J,J)}w_{l_{1}}w_{l_{2}}\\
        &\times\log_{2}\sum_{j=1}^{M}e^{-\SINR\cdot\|\mathbf{d}_{i,j}\|^{2}+2\sqrt{\SINR}\cdot\mathbb{R}\{(\chi_{l_{1}}\mathbf{\hat{i}}+\chi_{l_{2}}\mathbf{\hat{j}})\dotprod\mathbf{d}_{i,j}\}},
    \raisetag{5mm}
    \end{split}
    \label{eq:mi}
\end{equation}
where $\mathbf{d}_{i,j}$ is the distance vector between the $i^{th}$ and $j^{th}$ unitary power constellation coordinates, $\|\cdot\|$ is the Euclidean norm, $\mathbb{R}\{\cdot\}$ is the real part, $\dotprod$ is the dot product, and $\chi_{k}$ and $w_{k}$ are, respectively, the Gauss-Hermite quadrature nodes and weights, given for $J=5$ and $J=10$ in~\cite[Table 3.5.10 and Table 3.5.11]{olver2010nist}.

Figs.~\ref{fig:InformationRate}(a-b) show the impact of \gls*{mdg} on the average total pre-\gls*{fec} \gls*{ber} and on the  \gls*{mi}, respectively. We compare the analytical solutions given in~\eqref{eq:ber} and~\eqref{eq:mi}, with end-to-end simulations of a complete \gls*{sdm} systems. In comparison with the previous simulations,  which only simulate random channel matrices, we now consider the generation of random symbol sequences, their propagation along a randomly coupled channel, and the recovery processes at the receiver. In addition to the curves, we show the equivalent transparent reach distance as a function of \gls*{mdg}. As noted for the \gls*{sinr}, the information rate is overestimated at high \gls*{snr} and \gls*{mdg} values. These curves confirm the applicability of the derived analytical model to the design of practical \gls*{sdm} systems.

As we discussed earlier, the analysis is based on averages. Under practical scenarios, a certain operational margin is required to address the probability of outages. Nevertheless, as discussed in~\cite{arik2014diversity} and~\cite{ho2011frequency}, modal and frequency diversity induces an averaging effect that largely reduces the stochastic variation of capacity metrics.

\section{Performance Monitoring}
\label{sec:parameter}

This section addresses the task of performance monitoring in \gls*{sdm} receivers using information available in \gls*{mmse} equalizers. In particular, two metrics are estimated: the accumulated link \gls*{mdg} and the pre-equalizer \gls*{snr}. Indeed, although estimating the \gls*{snr} is trivial in traditional polarization-multiplexed receivers, this task is complicated in \gls*{sdm} systems because of the performance losses generated by \gls*{mdg}. Aligned with previous works \cite{ospina2020dsp, van2020experimental, ospina2021neural, ospina2021mode, ospina2022mdg, ospina2023digital, zischler2024estimating}, we also use the \gls*{mmse} equalizer coefficients as inputs for the proposed estimation technique. In~\cite{ospina2020dsp,van2020experimental}, we derive an analytical solution for $\sigma_{\mdg}$ estimation based on an correction factor applied to the standard deviation of the equalizer inverse eigenvalues. However, the proposed technique shows limitations under high-\gls*{mdg} and low-\gls*{snr} due to the multivalued behavior of the equalizer inverse eigenvalues. In~\cite{ospina2021neural}, we propose a neural-network-based technique for $\sigma_{\mdg}$ and $\SNR$ estimation, discussed in further detail in~\cite{ospina2022mdg}. In~\cite{zischler2024estimating}, a similar neural network approach was proposed for capacity loss estimation. All these techniques are validated experimentally demonstrating significant accuracy. This section leverages our recent results on \gls*{mmse} equalizers to derive a look-up-table-based solution for the pre-filtering \gls*{snr} and $\sigma_{\mdg}$ estimation with suitable accuracy for a wide range of values. 
\begin{figure}[!t]
    \centering
    \hspace{-0.25cm}\includegraphics{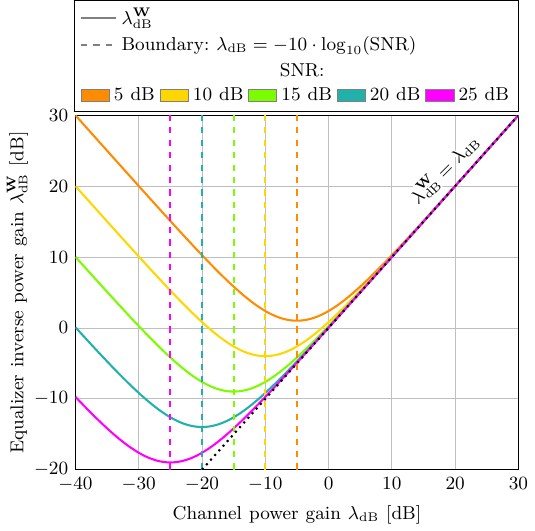}
    \caption{Relation between the channel logarithmic power gain, obtained from the eigenvalues of the channel matrix, and the equalizer inverse logarithmic power gain, obtained from the equalizer inverse matrix~\cite{ospina2020dsp}.}
\label{fig:Eigenvalues}
\end{figure}

\subsection{Statistics of the MIMO MMSE equalizer}

As the \gls*{mimo} \gls*{mmse} equalizer matrix $\mathbf{W}$ coincides with the zero-forcing solution $\mathbf{H}^{-1}$ in noise-free scenarios, the equivalent matrix product ${\mathbf{W}^{-1}(\mathbf{W}^{-1})^{H}}$ is often used for parameter estimation~\cite{ospina2020dsp, van2020experimental, ospina2021neural, ospina2021mode, ospina2022mdg, ospina2023digital}. The logarithmic scale \gls*{mimo} \gls*{mmse} eigenvalues $\lambda^{\mathbf{W}}_{\dB,i}$,  are related to the logarithmic scale power gains $\lambda_{\dB,i}$,  by ~\cite[Eq. (3)]{ospina2020dsp}
\begin{equation}
    \lambda^{\mathbf{W}}_{\dB,i}=20\cdot\log_{10}\left(\frac{1}{\SNR}10^{\frac{-\lambda_{\dB,i}}{10}}+1\right)+\lambda_{\dB,i}.
    \label{eq:lambdarelation}
\end{equation}

Fig.~\ref{fig:Eigenvalues} illustrates \eqref{eq:lambdarelation} for a wide range of \gls*{snr} values. In~\eqref{eq:lambdarelation} there is an inflection point at ${-10\cdot\log_{10}(\SNR)}$. For channel eigenvalues bellow $\lambda_{i}<\SNR^{-1}$, the linear scale equalizer inverse eigenvalues ($\lambda^{\mathbf{W}}_{i}$) is multivalued, impairing parameter estimation techniques~\cite{ospina2020dsp, van2020experimental, ospina2023digital}.

From~\eqref{eq:lambdarelation}, via transformation, we obtain the following relation between the \glspl*{pdf} of then logarithmic-scale eigenvalues $\lambda_{\dB}$ from the channel and the equalizer 
\begin{equation}
    \begin{gathered}
        f_{\lambda^{\mathbf{W}}_{\dB}}(\lambda^{\mathbf{W}}_{\dB}) =\frac{f_{\lambda_{\dB}}\left(g^{\bigoplus}(\lambda^{\mathbf{W}}_{\dB})\right)+f_{\lambda_{\dB}}\left(g^{\bigominus}(\lambda^{\mathbf{W}}_{\dB})\right)}{\sqrt{1-\frac{4}{\SNR}\cdot 10^{\frac{-\lambda^{\mathbf{W}}_{\dB}}{10}}}},\\
        \lambda^{\mathbf{W}}_{\dB}>10\cdot\log_{10}\left(\frac{4}{\SNR}\right),
    \end{gathered}
    \label{eq:flambdammse}
\end{equation}
where $g^{\bigoplus}(\cdot)$ and $g^{\bigominus}(\cdot)$ are, respectively, the positive and negative root solutions of
\begin{equation}
    \begin{split}
        g(\lambda^{\mathbf{W}}_{\dB})=10\cdot\log_{10}&\left[-\SNR^{-1}+ \frac{1}{2}10^{\frac{\lambda^{\mathbf{W}}_{\dB}}{10}}\right. \\
        &\left. \hspace{0.5em}\times\left(1\pm\sqrt{1-\frac{4}{\SNR}\cdot 10^{\frac{-\lambda^{\mathbf{W}}_{\dB}}{10}}}\right)\right].
    \end{split}
\end{equation}

The mean and standard deviation of the equalizer eigenvalues can be obtained via numeric integration of~\eqref{eq:flambdammse}.

\begin{figure}[!t]
    \centering
    \includegraphics{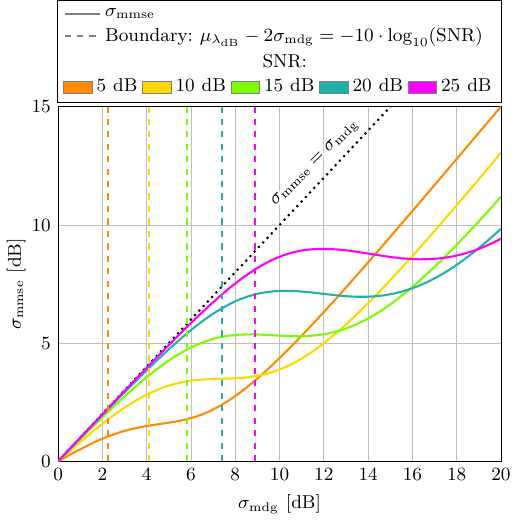}
    \caption{Relation between the standard deviation of the equalizer inverse logarithmic power gain ($\sigma_{\mmse}$) and the standard deviation of the channel logarithmic power gain ($\sigma_{\mdg}$) for $D=6$. The boundaries given by~\eqref{eq:limits} are also presented. The $\sigma_{\mmse}$ is obtained via numeric integration of~\eqref{eq:flambdammse}.}
    \label{fig:SigmaMmse}
\end{figure}

Fig.~\ref{fig:SigmaMmse} shows the relation between the standard deviation of the unlabeled eigenvalues derived from the channel matrix ($\sigma_{\mdg}$) and those derived from the equalizer inverse ($\sigma_{\mmse}$). Real and estimated values start to significantly deviate in the vicinity of the inflection point of \eqref{eq:lambdarelation}. From the boundaries of the limiting case given in~\eqref{eq:flambdadbwigner}, the equalizer inverse will have an approximate linear relation to the channel with sufficiently high \gls*{snr} if
\begin{equation}
    -2\sigma_{\mdg}+\mu_{\lambda_{\dB}}\gg-10\cdot\log_{10}(\SNR).
    \label{eq:limits}
\end{equation}
\begin{figure*}[!t]
    \centering
    \includegraphics{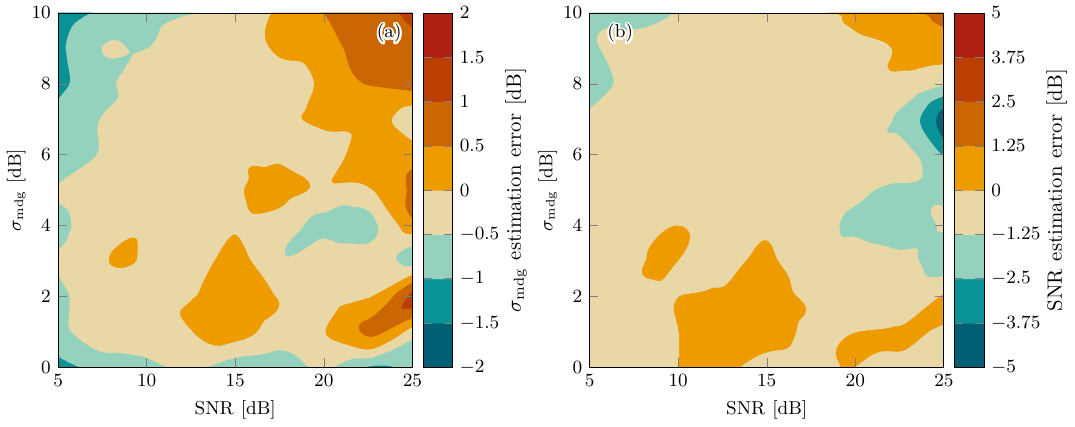}
    \caption{Parameter estimation error for $\sigma_{\mdg}$ (a) and for $\SNR$ (b) from~\eqref{eq:estimation}. The estimation technique utilizes a LUT, where the measured $\SINR$ and $\sigma_{\mmse}$ are compared against pre-calculated analytical results. Each estimated value is obtained from the average of 10 uncorrelated $\SINR$ and $\sigma_{\mmse}$ samples from a simulated setup with QPSK modulated symbol sequence.}
    \label{fig:SigmaSnrEstimation}
\end{figure*}

While we discussed before the asymptotic behavior of $\mathbf{W}^{-1}$ approaching $\mathbf{H}$ as \gls*{snr} tends to infinity,~\eqref{eq:limits} provides a rule of thumb for which the approximation ${\mathbf{W}^{-1}\approx\mathbf{H}}$ can be considered accurate for \gls*{mdg} estimation. Fig.~\ref{fig:SigmaMmse} visually illustrates how real and estimated eigenvalues (solid and dotted curves) deviate near the limit given by vertical dashed lines. 

Applying~\eqref{eq:flambdadb} to \eqref{eq:flambdammse} enables us to obtain the unlabeled $\lambda^{\mathbf{W}}_{\dB}$ distribution. As detailed in~\cite[Section~III]{zischler2025analytic}, the per-mode logarithmic gain \gls*{pdf} can be well approximated by a Gaussian distribution with mean $\mu_{\lambda_{\dB,i}}$ and standard deviation $\sigma_{\lambda_{\dB,i}}$. The per-mode $\lambda^{\mathbf{W}}_{\dB,i}$ \gls*{pdf}, in turn, can be obtained from the per-mode $\lambda_{\dB,i}$ \gls*{pdf} $f_{\lambda_{\dB,i}}(\lambda_{\dB,i})$ given by the approximate analytical solution in~\cite[Eq.~(7)]{zischler2025analytic}.

\subsection{Estimation of the pre-filtering SNR and MDG}

The $\sigma_{\mmse}$ and \gls*{sinr} are metrics can be readily estimated by receiver \gls*{dsp}. The \gls*{sinr} can be estimated using the error vector magnitude of the post-equalizer constellations. The $\sigma_{\mmse}$ can be estimated from the frequency domain equalizer inverse matrix. If the equalizer operates in the time domain, a discrete Fourier transform algorithm is required. Considering a discrete-time frequency domain equalizer $\mathbf{W}[\omega]$, $\omega\in[1, \Omega]$, where $\Omega$ is the equalizer discrete frequency bin count, $\sigma_{\mmse}$ can be estimated from
\begin{equation}
    \widehat{\sigma}_{\mmse}^{2}\hspace{-2pt}=\hspace{-6pt}\sum_{(\omega,i)=(1,1)}^{(\Omega,D)}\frac{\left[\text{eig}_{i}\left(\mathbf{W}^{-1}[\omega]\left(\mathbf{W}^{-1}[\omega]\right)^{H}\right)-\mu_{\mmse}\right]^{2}}{\Omega\cdot D-1},
\end{equation}
where $\text{eig}_{i}(\cdot)$ return the $i^{th}$ ordered eigenvalue, and $\mu_{\mmse}$ is the average of the equalizer inverse unlabeled eigenvalues, estimated from
\begin{equation}
    \mu_{\mmse}=\sum_{(\omega,i)=(1,1)}^{(\Omega,D)}\frac{\text{eig}_{i}\left(\mathbf{W}^{-1}[\omega]\left(\mathbf{W}^{-1}[\omega]\right)^{H}\right)}{\Omega\cdot D}.
\end{equation}

Interestingly, any unique \gls*{snr} and $\sigma_{\mdg}$ pair within a practical range of values yields a unique \gls*{sinr} and $\sigma_{\mmse}$ pair (shown numerically in \appref{app:b}). While a closed-form analytical equation relating \gls*{snr} and $\sigma_{\mdg}$ as functions of \gls*{sinr} and $\sigma_{\mmse}$ is unknown, it is possible to obtain this relation by numerically inverting the direct equation. Considering $\widehat{\SINR}$ and $\widehat{\sigma}_{\mmse}$ as  measured parameters, we calculate $\widehat{\SNR}$ and $\widehat{\sigma}_{\mdg}$ by minimizing the squared error to their analytical equivalents, i.e.,
\begin{equation}
    \begin{split}
        (\widehat{\SNR},\widehat{\sigma}_{\mdg})=&\text{argmin}_{\SNR,\sigma_{\mdg}}\left(\mathbb{E}\left\{\left[\widehat{\sigma}_{\mmse}-\sigma_{\mmse}\right]^{2}\right\}\right. \\
        &\left. +\mathbb{E}\left\{\left[\widehat{\SINR}-\SINR\right]^{2}\right\}\right),
    \end{split}
    \label{eq:estimation}
\end{equation}
where $\SINR$ and $\sigma_{\mmse}$ are obtained analytically from~\eqref{eq:sinrint} and~\eqref{eq:flambdammse} respectively.

The estimation process given in~\eqref{eq:estimation} can be performed iteratively, until a minimum error or iteration threshold is reached. Alternatively, given that the method relies on a numeric process, for fast real-time estimation, precomputed values can be stored in a \gls*{lut} with sufficient granularity.

Fig.~\ref{fig:SigmaSnrEstimation}  shows the $\sigma_{\mdg}$ and $\SNR$ estimation error from simulated values considering a complete \gls*{sdm} link transmission, with parameters as given by Table~\ref{tab:parameters}, and using a pre-computed \gls*{lut}. The used \gls*{lut} sweeps the \gls*{snr} for 100 logarithmic-spaced points from 0 to 30~dB, and the $\sigma_{\mdg}$ for 100 points from 0 to 20~dB.  Fig~\ref{fig:SigmaSnrEstimation}(a) confirms accurate $\sigma_{\mdg}$ estimation for the entire evaluated range, with some divergence at the lowest and highest \gls*{snr} values. Fig~\ref{fig:SigmaSnrEstimation}(b) shows discrepant values at high \gls*{snr} values, but accurate estimations for the rest of the evaluated region. The estimation error at high \gls*{snr} can be attributed to the previously mentioned underestimation of the analytical \gls*{sinr} within this region. In any case, practical scenarios with high SNR and high MDG are rare, as both effects accumulate with cascaded spans. From the estimated \gls*{snr} and $\sigma_{\mdg}$ values, additional meaningful performance metrics can be calculated, such as the \gls*{mdg}-induced capacity loss or effective \gls*{snr} loss~\cite{mello2020impact}.

\section{Conclusion}
\label{sec:conclusion}

We derive analytical expressions to quantify the effects of \gls*{mdg} in strongly-coupled \gls*{sdm} systems with \gls*{mimo} \gls*{mmse} equalization, including a closed-form expression for the system information rate. The presented solutions describe with suitable accuracy the average behavior of such systems within practical values of \gls*{snr} and \gls*{mdg}. The provided solutions enable fast system analysis, avoiding the use of costly semi-analytical simulations for average capacity evaluation. Furthermore, we review the statistical behavior of the  \gls*{mimo} \gls*{mmse} equalizer coefficients under strongly-coupled \gls*{sdm} systems. We show a practical region where the \gls*{mimo} \gls*{mmse} equalizer inverse can be utilized as an approximation to the channel. We also demonstrate that the mapping between relevant parameters before and after the MMSE equalizer are single-valued, enabling a look-up-table based method for performance monitoring.

\appendices
\def\sectionautorefname{appendix}

\section{Analytical Solution for the Average SINR of a System with MDG and Using an MMSE Equalizer}
\label{app:a}

Considering the eigenvector matrix $\mathbf{Q}$ and the eigenvalue diagonal matrix $\boldsymbol{\lambda}$ from the eigendecomposition of $\mathbf{H}^{H}\mathbf{H}$, we can rewrite~\eqref{eq:sinr} as
\begin{equation}
    \begin{split}
        \SINR_{i}&=\frac{1}{\left[\left(\mathbf{Q}\mathbf{Q}^{-1}+\SNR\cdot\mathbf{Q}\boldsymbol{\Lambda}\mathbf{Q}^{-1}\right)^{-1}\right]_{i,i}}-1\\
                 &=\frac{1}{\left[\mathbf{Q}\left(\mathbf{I}+\SNR\cdot\boldsymbol{\Lambda}\right)^{-1}\mathbf{Q}^{-1}\right]_{i,i}}-1.
    \end{split}
    \label{eq:sinr1}
\end{equation}

The eigenvalues $\boldsymbol{\lambda}$ are the same values as obtained from the eigendecomposition of $\mathbf{H}\mathbf{H}^{H}$, given that the channel matrix is square. However, the eigenvectors have distinct values. Nevertheless, $\mathbf{H}^{H}\mathbf{H}$ is also a valid \gls*{gue} matrix in a logarithmic scale.

The eigenvector matrix $\mathbf{Q}$ is given by
\begin{equation}
    \mathbf{Q}=\begin{bmatrix}
        v_{1,1} & v_{2,1} & \cdots & v_{D,1} \\
        v_{1,2} & v_{2,2} & \cdots & v_{D,2} \\
        \vdots  & \vdots  & \ddots & \vdots  \\
        v_{1,D} & v_{2,D} & \cdots & v_{D,D} \\
    \end{bmatrix},
\end{equation}
where $v_{i,j}$ is the $j^{th}$ element of the eigenvector $\mathbf{v}_{i}$. Given that $\mathbf{Q}^{-1}=\mathbf{Q}^{H}$, the element $(i,i)$ of the denominator of~\eqref{eq:sinr1} can be simplified to scalar operations of the eigenvector elements
\begin{equation}
    \SINR_{i} = \left[\sum_{j=1}^{D}|v_{i,j}|^{2}\left(1+\SNR\cdot\lambda_{j}\right)^{-1}\right]^{-1} - 1.
    \label{eq:sinr2}
\end{equation}

The eigenvector ensemble is uniformly distributed along the set of unitary norm vectors~\cite{mehta2004random, o2016eigenvectors}. For any eigenvector $\mathbf{v}_{i}$, its elements are distributed such that the norm is unitary, resulting in the following joint \gls*{pdf}
\begin{equation}
    f_{\mathbf{v}_{i}}(\mathbf{v}_{i})=\alpha_{\mathbf{v}_{i},D}\delta\left(1-\sum_{j=1}^{D}|v_{i,j}|^{2}\right),~0\le|v_{i,j}|^{2}\le 1,
\end{equation}
where $\delta(\cdot)$ is the Kronecker delta, and $\alpha_{\mathbf{v}_{i},D}$ is a normalization factor obtained from the surface area of the $D^{th}$ dimensional unit sphere
\begin{equation}
    \alpha_{\mathbf{v}_{i},D}^{-1}=\frac{2^{1-D}\pi^{\frac{D}{2}}}{\Gamma\left(\frac{D}{2}\right)},
\end{equation}
where $\Gamma(\cdot)$ is the gamma function.

Each eigenvector element squared modulus has the following distribution~\cite{livan2018introduction}
\begin{equation}
    f_{|v_{i,j}|^{2}}(|v_{i,j}|^{2})=(D-1)(1-|v_{i,j}|^{2})^{D-2},
\end{equation}
with mean
\begin{equation}
    \mathbb{E}\left\{|v_{i,j}|^{2}\right\}=\frac{1}{D}.
    \label{eq:eigenvectormean}
\end{equation}

We can rewrite the term inside the brackets in~\eqref{eq:sinr2} as
\begin{equation}
    \SINR_{i} = \left[D\sum_{j=1}^{D}\frac{1}{D}|v_{i,j}|^{2}\left(1+\SNR\cdot\lambda_{j}\right)^{-1}\right]^{-1} - 1,
\end{equation}
where the summation in respect to $j$ is a sample mean. The sample mean converges to the expectation as $D$ approaches infinity
\begin{equation}
    \lim_{D\rightarrow\infty}\SINR_{i} = \hspace{-2mm}\lim_{D\rightarrow\infty}\left[D\cdot \mathbb{E}\left\{|v_{i,j}|^{2}\left(1+\SNR\cdot\lambda\right)^{-1}\right\}\right]^{-1}\hspace{-2mm}- 1.
\end{equation}

We omit $j$ of $\lambda$ to avoid confusion, as the expectation of the marginal eigenvalues is not equal to that of the unlabeled distribution due to the previously defined eigenvalue ordering. Expectation is carried out over the set of eigenvector elements  $\mathbf{v}_{i}$ and the set of eigenvalues $\lambda$. As the \gls*{gue} is invariant under unitary transformations~\cite{dyson1962threefold, mehta2004random}, the eigenvectors and eigenvalues are uncorrelated. Therefore, we can split the expectation terms as
\begin{equation}
    \begin{split}
        \lim_{D\rightarrow\infty}&\SINR_{i} =  \\ 
        \lim_{D\rightarrow\infty}&\left[D\cdot \mathbb{E}\left\{|v_{i,j}|^{2}\right\}\mathbb{E}\left\{\left(1+\SNR\cdot\lambda\right)^{-1}\right\}\right]^{-1} - 1.
    \end{split}
\end{equation}

We can finally simplify the equation with~\eqref{eq:eigenvectormean} and omit the \gls*{sinr} index, as
\begin{equation}
    \begin{split}
        \lim_{D\rightarrow\infty}\SINR = \left[\mathbb{E}\left\{\left(1+\SNR\cdot\lambda\right)^{-1}\right\}\right]^{-1} - 1.
    \end{split}
    \label{eq:sinrexpectation}
\end{equation}
Using random variable transformation, and expanding the expectation operator, yields
\begin{equation}
    \SINR = \left[\int_{x_{0}}^{x_{f}}\frac{10\cdot f_{\lambda_{\dB}}\left(10\cdot\log_{10}\left(\frac{1-x}{\SNR\cdot x}\right)\right)}{\ln(10)(1-x)}dx\right]^{-1} - 1.
    \label{eq:a0}
\end{equation}
where the integration limits $x_{0}$ and $x_{f}$ are restricted by the term inside the logarithm
\begin{equation}
    \frac{1-x}{\SNR\cdot x}>0,
\end{equation}
resulting in
\begin{equation}
    0<x<1.
\end{equation}

Replacing the limits in~\eqref{eq:a0} we obtain~\eqref{eq:sinrint}.

\section{Proof of ($\SINR$,$\sigma_{\mmse}$) and ($\SNR$,$\sigma_{\mdg}$) single-valuedness}
\label{app:b}

The inverse function theorem states that any continuously differentiable function will be invertible in a region around a certain point if the Jacobian determinant at that point is non-zero~\cite{clarke1976inverse}. For our purposes, the invertibility of a ($\SINR$,$\sigma_{\mmse}$ ) pair as a function of a ($\SNR$,$\sigma_{\mdg}$) pair, within a region of interest, guarantees that there is a unique single-valued mapping between both pairs.

The values of $\SINR$ and $\sigma_{\mmse}$ are smooth functions for all $\SNR$ and $\sigma_{\mdg}$ values of interest (${\SNR>0~\dB}$ and ${\sigma_{\mdg}>0~\dB}$). The $\SINR$ is monotonic decreasing in respect to $\sigma_{\mdg}$ and increasing in respect to $\SNR$ within the region of interest. As $\sigma_{\mmse}$ is a standard deviation value, it is real, finite, and non-negative for any valid ${f_{\lambda_{\dB}^{\mathbf{W}}}(\lambda_{\dB}^{\mathbf{W}})}$ \gls*{pdf}.

The Jacobian matrix for $\SINR$ and $\sigma_{\mmse}$ is given by
\begin{equation}
    \mathbf{J}=
    \begin{bmatrix}
        \dfrac{\partial \SINR}{\partial \SNR} & \dfrac{\partial \SINR}{\partial \sigma_{\mdg}}\\[1em]
        \dfrac{\partial \sigma_{\mmse}}{\partial \SNR} & \dfrac{\partial \sigma_{\mmse}}{\partial \sigma_{\mdg}}
    \end{bmatrix}
    \label{eq:J}
\end{equation}

\begin{figure}[!t]
    \centering
    \includegraphics{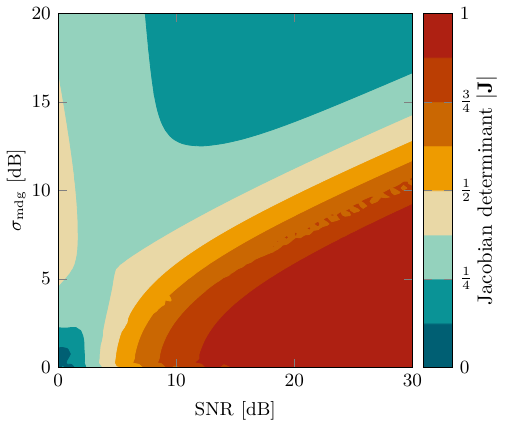}
    \caption{Jacobian determinant of ($\SINR$, $\sigma_{\mmse}$) with respect to ($\SNR$, $\sigma_{\mdg}$). The solutions were obtained numerically, from the analytical solutions for the limiting case ($D\rightarrow\infty$).}
    \label{fig:Jacobian}
\end{figure}

Obtaining a closed-form solution for~\eqref{eq:J} is not trivial. However, it can be solved numerically within a desired range. In Fig.~\ref{fig:Jacobian}, we evaluate the Jacobian determinant along a wide range of values (${\SNR\in(0,30)~\dB}$, ${\sigma_{\mdg}\in(0,20)~\dB}$). We evaluate both \gls*{snr} and \gls*{sinr} in logarithmic scale, as it is a monotonic function that preserves single-valuedness while keeping the unit scale within practical ranges. Fig.~\ref{fig:Jacobian} shows that, with the asymptotic exception of ${\SNR\rightarrow 0~\dB}$ and ${\sigma_{\mdg}\rightarrow 0~\dB}$, $|\mathbf{J}|$ is never zero. Therefore, any  ($\SINR$,$\sigma_{\mmse}$) pair values is related to a single unique ($\SNR$, $\sigma_{\mdg}$) pair within the evaluated region.

The Jacobian determinant also provides a measure of the intensity of correlation between sets, i.e. how a fixed deviation along the input variables propagates to the outputs. The greater the absolute Jacobian determinant, the greater the propagated deviation. Fig.~\ref{fig:Jacobian} shows that the highest absolute Jacobian determinant, and therefore, the higher correlation between the measured and desired variables, is at high \gls*{snr} and low \gls*{mdg} values. We also derived this conclusion previously in Section~\ref{sec:power}. At low \gls*{snr} and low \gls*{mdg} and at high \gls*{snr} and high \gls*{mdg} the absolute determinant is small. Therefore, in these regions,  measurements of $\SINR$ and $\sigma_{\mmse}$ with higher precision are required for proper $\SNR$ and $\sigma_{\mdg}$ 
 estimation.


\bibliographystyle{IEEEtran}
\bibliography{references}

\end{document}